# Topological Wannier cycles for the bulk and edges

Ze-Lin Kong[1], Zhi-Kang Lin[1, †] and Jian-Hua Jiang[1, 2, †]

[1]*School of Physical Science and Technology, & Collaborative Innovation Center of Suzhou Nano Science and Technology, Soochow University, 1 Shizi Street, Suzhou, 215006, China*

[2]*Key Lab of Advanced Optical Manufacturing Technologies of Jiangsu Province & Key Lab of Modern Optical Technologies of Ministry of Education, Soochow University, Suzhou 215006, China.*

[†]To whom correspondence should be addressed: linzhikangfeynman@163.com (ZKL) and jianhuajiang@suda.edu.cn (JHJ).

## Abstract

**Topological materials are often characterized by unique edge states which are in turn used to detect different topological phases in experiments. Recently, with the discovery of various higher-order topological insulators, such spectral topological characteristics are extended from edge states to corner states. However, the chiral symmetry protecting the corner states is often broken in genuine materials, leading to vulnerable corner states even when the higher-order topological numbers remain quantized and invariant. Here, we show that a local artificial gauge flux can serve as a robust probe of the Wannier type higher-order topological insulators which is effective even when the chiral symmetry is broken. The resultant observable signature is the emergence of the cyclic spectral flows traversing one or multiple band gaps. These spectral flows are associated with the local modes bound to the artificial gauge flux. This phenomenon is essentially due to the cyclic transformation of the Wannier orbitals when the local gauge flux acts on them. We extend topological Wannier cycles to systems with $C_2$ and $C_3$ symmetries and show that they can probe both the bulk and the edge Wannier centers, yielding rich topological phenomena.**

## Introduction

The discovery of higher-order topological insulators (HOTIs) generalizes the celebrated bulk-edge correspondence [1, 2] to various multidimensional bulk-boundary correspondences [3-11], such as the bulk-edge-corner correspondence. For instance, in two-dimensional (2D) HOTIs, the emergence of zero-dimensional (0D) corner states in the bulk band gap is often used as a spectral signature of the multidimensional topological physics and the higher-order topology. Such a phenomenon, though convenient for experimental measurements [12-31], however, has deficiencies. Due to chiral symmetry breaking in most materials, the frequencies of the corner states are not pinned to the middle of the bulk band gap. Under certain situations, these corner states can even be tuned out of the bulk band gap and merge into the bulk bands [32]. In these situations, the spectral probe of the higher-order topology becomes impossible, making the detection of the higher-order topology relying on the detection of the fractional corner charges which are much more difficult to measure in experiments [11, 33, 34].

An alternative approach is to use a highly localized gauge flux as a tool to probe higher-order topological insulating phases with filling anomaly [11]. As demonstrated in Ref. [34] for HOTIs with four-fold rotation symmetry ($C_4$), the local gauge flux induces topological Wannier cycles in HOTIs that are Wannier representable (also called as the obstructed atomic insulators [35-38]). This emergent effect, which is the focus of this paper, is due to the cyclic evolution among the Wannier orbitals at the same Wannier center but of different spatial symmetries. Topological Wannier cycles are manifested as the cyclic spectral flows that traverse multiple band gaps when the local gauge flux encloses the Wannier center. These spectral flows evolve from the bulk states at $\Phi = 0$ and $\Phi = 2\pi$ (In this paper, we omit the unit of the gauge flux $\hbar/e$), and gradually become localized modes bound to the gauge flux within the bulk band gaps [34]. However, it is unclear whether topological Wannier cycles can be extended to HOTIs of different spatial symmetries or even be generalized to probe the topology of edge states.

Here, we first generalize topological Wannier cycles to three-fold ($C_3$) and two-fold ($C_2$) rotation symmetric HOTIs. In the $C_3$-symmetric HOTIs, we show that the topological Wannier cycles are developed from the bulk and has no effect on the edge or corner states, confirming that topological Wannier cycles can serve as a probe of HOTIs with other symmetries. Here, the artificial local gauge flux is invoked by consecutive procedures involving the dimension extension, introducing a step screw dislocation (SSD), and dimensional reduction [34]. Quite differently, for $C_2$ symmetric HOTIs, the artificial local gauge flux is invoked by introducing an zigzag edge boundary, leading to a novel mechanism for topological edge states in two-dimensional (2D) HOTIs. Furthermore, we discover that in finite $C_4$ symmetric HOTI systems, both the bulk and the edge states can lead to topological Wannier cycles yet of distinctive nature.

## HOTIs with various spatial symmetry

The topological Wannier cycles in HOTIs are previously elaborated in the 2D Su-Schrieffer-Heeger (SSH) model with $C_4$ symmetry using designed sonic crystals [34]. Here, without loss of generality, we revisit another two prototype models that are defined for spinless particles on lattices with $C_3$ or $C_2$ symmetry. These tight-binding (TB) models are introduced in Fig. 1, where, a breathing kagome model realizes the HOTI with $C_3$ rotation symmetry [see Fig. 1(a)]. As a special case, we also consider the SSH model as the simplest model with $C_2$ symmetry. Here, the SSH model is constructed by weakly coupling the one-dimensional (1D) SSH chain along the vertical direction [see Fig. 1(c)]. Although such a SSH model should not be classified as a HOTI, it has well-defined Wannier centers and filling anomaly which can thus serve as a particularly simple case to illustrate the topological Wannier cycles.

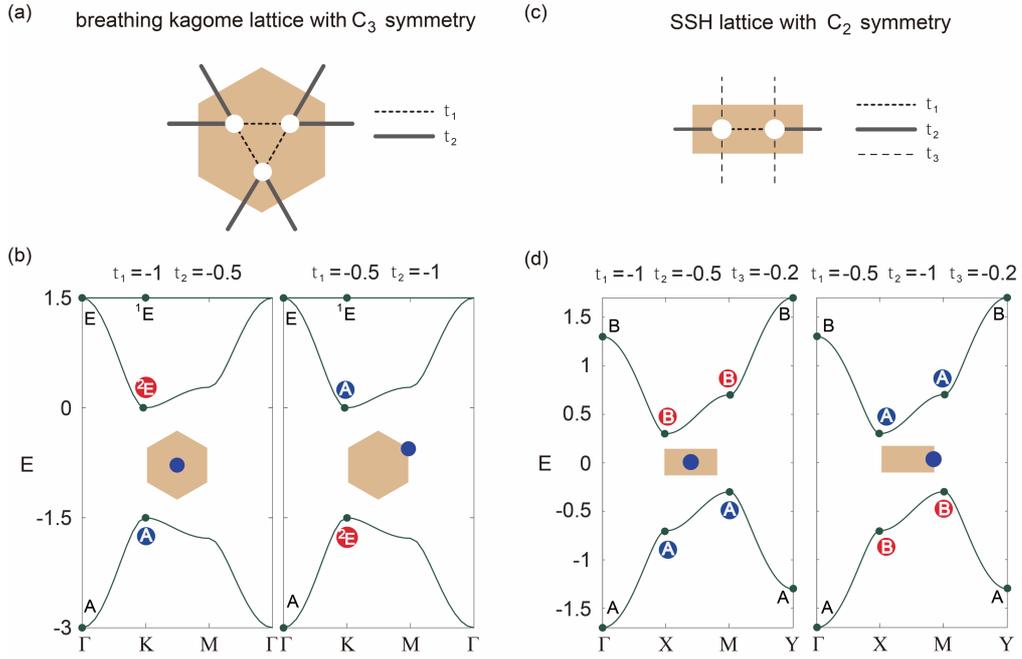

Fig. 1 | (Color online) (a) and (b) Schematic illustrations of the 2D breathing kagome lattice with $C_3$ symmetry and the SSH model with $C_2$ symmetry. (c) and (d) Band structures and symmetry properties of the bands for the TB models in (a) and (b), separately. For each figure, the left panel indicates the trivial phase while the right panel indicates the topological phase. Insets illustrate the location of the Wannier centers in a unit-cell. Symbols represent the little group representations at the high-symmetry points of the Brillouin zone. Symbols in colored circles are used to highlight the different symmetry properties between the trivial and topological phases. The TB parameters are listed above each panel.

We first give the bulk bands and their properties for these TB models in both the topological and the trivial phases. The Wannier orbitals and Wannier centers in real space of these energy bands are also illustrated, respectively, in Fig. 2 and insets of Figs. 1(b) and 1(d), which can be deduced from the theory of elementary band representations (see Supplementary Material). As shown in Fig. 1(b), the breathing kagome model has three bands separated by one energy gap between the 1st and the 2nd band. These energy bands are composed of Wannier orbitals with different $C_3$-rotation eigenvalues, $g_n^{(3)} = e^{\frac{in2\pi}{3}}$ ($n = 0, 1, 2$) [see Fig. 2(a)]. In the trivial phase, the Wannier centers of all bands locate at the unit-cell center. In contrast, in the topological phase, the Wannier centers locate at the corners of the unit-cell. The trivial phase is realized when the intra-unit-cell coupling $t_1$ is stronger than the inter-unit-cell coupling $t_2$, while the topological phase is realized when the inter-unit-cell coupling $t_2$ is stronger than the intra-unit-cell coupling $t_1$.

Similarly, for the SSH model, if the inter-unit-cell coupling $t_2$ is stronger than the intra-unit-cell coupling $t_1$, the system is in the topological phase and the Wannier centers are at the edge of the unit-cell. Reversing the inter-unit-cell and the intra-unit-cell couplings leads to the trivial phase where the Wannier centers are at the unit-cell center. There are two bands separated by one energy gap, which originate from two Wannier orbitals with different $C_2$-rotation eigenvalues, $g_n^{(2)} = e^{\frac{in2\pi}{2}}$ ($n = 0, 1$) [see Fig. 2(b)]. The topological indexes for both the breathing kagome and the SSH models, based on the symmetry eigenvalues at the high-symmetry points (HSPs) of the Brillouin zone, are presented in Supplementary Material.

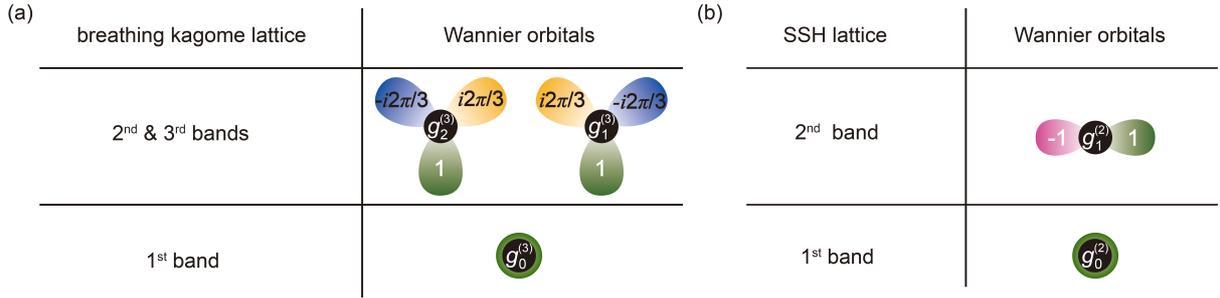

Fig. 2 | (Color online) Wannier orbitals of energy bands in the breathing kagome lattice model (a) and the SSH model (b). The phase differences between adjacent petals indicate the rotation symmetries $g_n^{(3)} = e^{\frac{in2\pi}{3}}$ with $n = 0, 1, 2$ in (a) and $g_n^{(2)} = e^{\frac{in2\pi}{2}}$ with $n = 0, 1$ in (b).

## Local artificial gauge flux

We now create the local artificial gauge flux through three consecutive procedures which include the dimension extension, creating a SSD, and dimensional reduction [34] (see Fig. S1 in Supplementary Material). After the dimension extension and the creation of a dislocation, the system of the 2D breathing kagome model has $3N(3N - 1)/2$ unit-cells (with $N = 10$ in our calculations) in the $x$-$y$ plane where the dislocation core is at the center of this plane [see Fig. 3(a)]. Therefore, each sector has $N(3N - 1)/2$ unit-cells. Such a designed system has an emergent screw symmetry $S_{3z} = C_3 * \mathcal{L}_z(\frac{1}{3})$ with $\mathcal{L}_z(\frac{1}{3}) := z \to z + \frac{1}{3}$ (we set the lattice constants along all directions as unit for the TB models in this work) which play a pivotal role in emergent physics.

For the dimension-extended SSH model, we can create a step glide dislocation that is finite in the $x$-direction and periodic in the $y$- and $z$-directions [see Fig. 3(b)]. The system then has an emergent glide symmetry $G_{2z} = C_2 * \mathcal{L}_z(\frac{1}{2})$ with $C_2 \coloneqq x \to -x$. However, such a structure has translation symmetry in the $y$ direction. The topological physics emerges essentially in the $x$-$z$ plane where the step glide dislocation becomes a zigzag edge boundary.

We now employ the dimensional reduction procedure. Dimensional reduction [53] is a tool to map higher-dimensional physics into lower-dimensional physics which is widely used in the study of topological physics. Here, this procedure is achieved by the Fourier transformation along the $z$-direction. After the Fourier transformation, the dimension along the $z$-direction becomes a system with parameter $k_z$ and thus leads to the reduction of dimension by one.

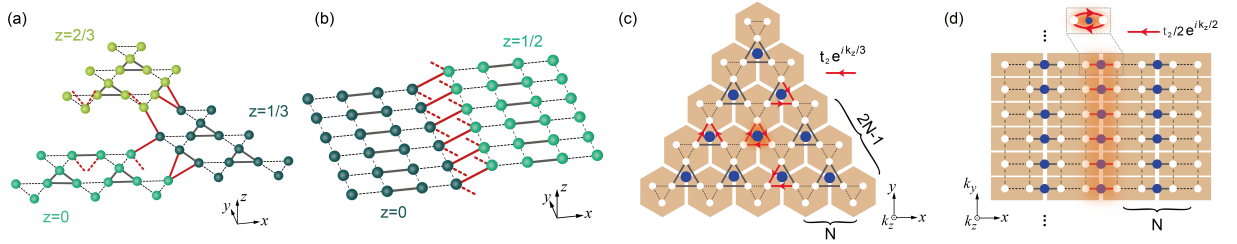

Fig. 3 | (Color online) (a) and (b) Realization of the single-plaquette artificial gauge flux insertion in the 2D breathing kagome model (a) and the SSH model (b) via three consecutive procedures. The structures are periodic in the $z$-direction while only the primary parts are shown. The coupling denoted by dashed red lines indicates the periodicity. The resultant effective models after these procedures are presented in (c) and (d), separately. The artificial gauge fields are reflected by the couplings with arrows: hopping along the arrow picks up a gauge phase term of $e^{i\Phi/3}$ or $e^{i\Phi/2}$ with $\Phi = k_z$.

The effective models after the three consecutive procedures are given in Figs. 3(c) and (d). Due to the Fourier transformation along the $z$-direction, each tilted coupling between the adjacent sectors will pick up a gauge phase factor $e^{i\theta}$ along the arrows. This gauge phase is $\theta = \frac{k_z}{3}$ or $\frac{k_z}{2}$, separately for the 2D breathing kagome model and the SSH model. With the gauge phase configurations illustrated in Figs. 3(c) and (d), the highly localized artificial gauge flux confined in a single-plaquette, $\Phi = k_z \in [0, 2\pi]$, emerge at the center of the system for the breathing kagome model. For the SSH model, the localized artificial gauge flux can be regarded as to be confined in the center of each SSH chain [see the top panel in Fig. 3(d)].

# Topological Wannier cycles induced by the local gauge flux

We now show that the single-plaquette artificial gauge flux can lead to the topological Wannier cycles. The key here is to understand the cyclic transformation of the Wannier orbitals under the local gauge flux insertion. For instance, for the breathing kagome model, the three bands are made of three Wannier orbitals with the $C_3$-rotation eigenvalues, $g_n^{(3)} = e^{\frac{in2\pi}{3}}$ ($n = 0, 1, 2$). In the topological phase, the artificial gauge flux $\Phi$ acts only on the Wannier orbitals in the central plaquette [see the orange zone in Fig. 3(c)], then these Wannier orbitals now have the $C_3$-rotation eigenvalues, $e^{i[n\left(\frac{2\pi}{3}\right)+\frac{\Phi}{3}]}$. This change can also be understood by replacing the $C_3$ rotation to the $S_{3z}$ rotation in the structure with a step screw dislocation. Since $S_{3z} = C_3 * \mathcal{L}_z(\frac{1}{3})$, the $C_3$-eigenstates at $k_z$ pick up an additional phase, $e^{ik_z/3} = e^{i\Phi/3}$, due to the translation along the $z$-direction $\mathcal{L}_z(\frac{1}{3})$. When a $\Phi = 2\pi$ gauge flux is inserted in the central plaquette, the rotation eigenvalues transform as, $g_n^{(3)} \to g_{n+1}^{(3)}$. Because $n = 0, 1, 2$ is of modulo 3 quantity, such transformation forms a closed cycle, $g_0^{(3)} \to g_1^{(3)} \to g_2^{(3)} \to g_0^{(3)}$. Since each Wannier orbital comprises and represents a bulk band, the cyclic transformation among the Wannier orbitals leads to a cyclic spectral flow among the bulk bands, which is exactly the topological Wannier cycle, as shown in Fig. 4(a). In contrast, if the system is in the trivial phase with $|t_1| > |t_2|$, then the Wannier orbitals live on the plaquettes that do not carry any gauge flux, and the inserted gauge flux thus does not affect the spectrum.

The Aharonov-Bohm principle [54] states that essentially the gauge flux induces a change in the phase of the wavefunctions. Normally, inserting a gauge flux $\Phi = 2\pi$ has no physical effect, since the phase $2\pi$ is equivalent to phase $0$. However, in topological systems, inserting a gauge flux $\Phi = 2\pi$ can lead to nontrivial effects. The topological Wannier cycle is one of the nontrivial effects that can emerge in topological crystalline systems.

We now analyze the filling anomaly of the breathing kagome model. Consider a triangular, finite 2D system with $3N(3N-1)/2$ unit-cells without the artificial gauge flux. Such a finite system has the $C_3$ rotation symmetry, while the lattice translation symmetry is broken. In the trivial case, because the Wannier orbitals are at the unit-cell center, there are $3N(3N-1)/2$ bulk states in each bulk band.

In the topological case, the filling anomaly emerges, the situation is quite different when the Wannier centers are exposed at edges and corners. We find that there are $9(N-1)$ edge states in the band gap [see the green region in Fig. 4(a)]. In addition, there are two sets of corner states whose total energy is pinned to zero due to the chiral symmetry. However, only one set is in the band gap in our calculation [see the blue line in Fig. 4(a)]. Most importantly, there are $\frac{9N(N-1)}{2} + 1$ bulk states in the first bulk band. The number of bulk states is equal to the number of Wannier centers in the bulk region. Moreover, this number is also the consequence of the band topology and has several highly nontrivial features. First, it predicts a fractional corner charge. Since the finite system is $C_3$-symmetric, each third sector then has an electron charge of $\left(\frac{3N(N-1)}{2} + \frac{1}{3}\right)e$ when the first bulk band is filled. The fractional part of this charge, $e/3$ or $-2e/3$, is consistent with the fractional charge at the corner unit-cell, as shown in Fig. 4(c). We remark that the corner and edge charges calculated from the first-principle method (see Supplementary Material) are approximate to the values labeled in Figs. 4(c). The small deviations are due to the spatial extensions of the Wannier orbitals and the finite size effect. Nevertheless, the fractional charge $e/3$ (or $-2e/3$), are always quantized for each sector due to the $C_3$ symmetry. Second, the number of bulk states also dictates the topological Wannier cycles. Because when a $\Phi = 2\pi$ gauge flux is inserted into the central plaquette, the eigenstates transform cyclically. This cyclic transformation must be completed within a group of three eigenstates. However, the number of bulk states in the first bulk band, $\frac{9N(N-1)}{2} + 1$, is *not* an integer multiple of three. Simultaneously, there are two bulk states above the gap which cannot find their partner to form a group as well. The cyclic transformation of these eigenstates cannot be fulfilled within their bulk continua. The only way to fulfill the gauge invariance, is to have cyclic transformation among themselves through in-gap modes bound to the gauge flux which is exactly the emergent topological Wannier cycles shown in Fig. 4(a) as calculated from the TB model in Fig. 3(c). It is seen that the edge and corner states appear in the band gap, yet they do not evolve with the inserted gauge flux and have nothing to do with the spectral flows and the topological Wannier cycles. We emphasize that since the numbers of the edge and corner states are both integer multiples of three, whether these states are filled or not does not change the above reasoning. Therefore, the topological Wannier cycles revealed here is purely a bulk effect and extends the idea in Ref. [34] to other spatial symmetries. In real space, this effect is manifested as the emergence of the topological modes bound to the SSD in a 3D topological system made of stacking the 2D breathing kagome lattice. We remark that although this phenomenon is seemingly like the helical 1D topological states bound to screw dislocation in weak topological insulators [43, 44], the underlying physics is completely different since the scenario revealed here is based on spinless models (with Wannier orbitals and crystalline symmetries) instead of the spin-1/2 models (with spin-orbital couplings and time-reversal symmetry) studied in Refs. [43, 44].

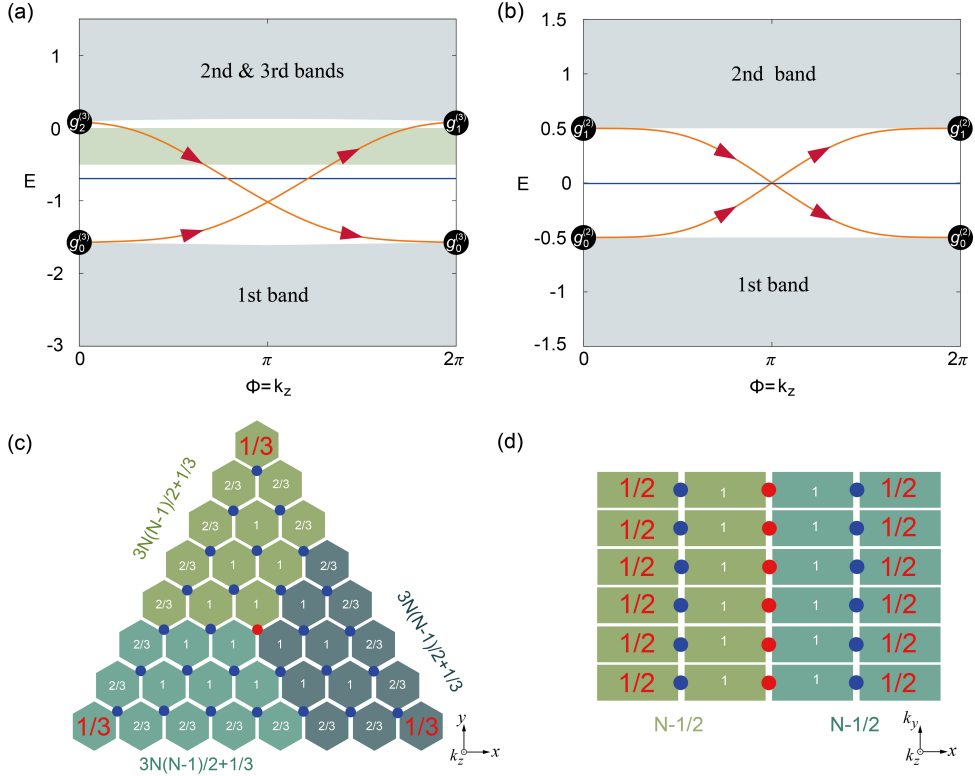

Fig. 4 | (Color online) Topological Wannier cycles and fractional charges in the breathing kagome and SSH models. (a) and (b): The energy spectra of the TB models as functions of the artificial gauge flux $\Phi = k_z$ for the finite systems in Fig. 3(c) and in Fig. 3(d). The TB parameters are $t_1 = -0.5$, $t_2 = -1$ and N = 10 in both two cases. Here, the energy spectra in (b) is calculated for the SSH model which is finite in the $x$-direction and periodic in the $y$-direction with a vertical coupling $t_3 = -0.2$ and $k_y = 0.5\pi$. The grey and green regions denote the bulk and edge states, respectively. The orange curves with red arrows represent the spectral flows between the bulk bands. The blue lines in (a) and (b) denote the corner states of the breathing kagome model and edge states of the SSH chain. Both the edge and corner states have nothing to do with the spectral flows across the band gaps, i.e., the topological Wannier cycles. (c) and (d): Distribution of the electron charges. For the breathing kagome and the SSH models, each sector has $\frac{3N(N-1)}{2} + \frac{1}{3}$ and $N - \frac{1}{2}$ electron charges, respectively, which leads to the fractional charges 1/3 or 1/2 at the corner- or edge-unit-cells. The blue dots denote the Wannier centers located in the bulk region, wherein, the central one is marked with red color.

A closer examination of the spectral flows indicates that at $\Phi = 0$ the system returns to the breathing kagome model. The spectral flows start from the bulk states in the $\Phi = 0$ limit and evolve gradually into the localized states bound to the artificial gauge flux at the central plaquette. At $\Phi = \pi$, the two time-reversal symmetric topological boundary states become degenerate since the $C_3$-rotation eigenvalues of the topological boundary states are mutually conjugated. At $\Phi = 2\pi$, the spectral flow ends at a bulk band that is different from the original bulk band it starts from. The cyclical spectral flows perfectly satisfy the periodicity as $\Phi = k_z$ goes from 0 to $2\pi$, yet with the connections determined by the bulk invariants (see analysis below), they manifest the nontrivial topological spectral evolutions across the band gaps.

The above analysis shows that the number of eigenstates in each bulk band is a crucial indicator of the higher-order band topology. More precisely, the number of eigenstates modulo three for all the bulk bands below the band gap of concern is the true indicator. We now connect this indicator with the bulk topological invariants which are called the real-space topological invariants (RSTIs). The theory of RSTIs is proposed and developed in Refs. [10, 39, 40]. The RSTIs are good quantum numbers in real space that connect to the symmetry indicators of the Bloch bands [35-38] and the topological invariants (such as the Chern and $Z_2$ numbers) related to the symmetry indicators [34]. For a finite $C_3$-symmetric 2D system, the eigenstates have three irreducible representations (IRs), i.e., $A$, $^2E$, and $^1E$, corresponding to the three $C_3$-rotation eigenvalues $g_n^{(3)} = e^{in(\frac{2\pi}{3})}$ with $n = 0$, 1 and 2 mod 3, separately. The RSTIs are defined as the following three integer invariants

$$(\delta_1, \delta_2) = (m(g_2) - m(g_0), m(g_1) - m(g_0)). \qquad (1)$$

Here, $\delta_2$ is equal to $\delta_1$ in the presence of time-reversal symmetry. $m(g_n)$ denotes the multiplicity of the eigenstates with $C_3$-rotation eigenvalue $g_n$ for the finite system. Interestingly, although the above definition is based on finite systems, the RSTIs can be deduced from the symmetry indicators of the Bloch bands below the band gap of concern at the HSPs in the Brillouin zone [39]. In our $C_3$-symmetric system, the RSTIs are calculated as follows [39],

$$\delta_1 = \frac{2}{3}m(^2E_\Gamma) + \frac{1}{3}m(^1E_\Gamma) - m(A_K) - m(^2E_K),$$

$$\delta_2 = \frac{1}{3}m(^2E_\Gamma) + \frac{2}{3}m(^1E_\Gamma) - m(^2E_K). \qquad (2)$$

Here, the symbols in the bracket represent the IRs at the HSPs in the Brillouin zone (the HSPs are the $\Gamma$, $M$, and $K$ points which are represented by the subscripts). The $m$'s denote the multiplicity of the IRs below the band gap. For instance, $m(A_K)$ denotes the multiplicity of the IR $A$ at the $K$ point. Based on the IRs labeled at the HSPs in Fig. 1c, the RSTIs of the band gap are calculated as $(-1, -1)$ and $(0, 0)$ for the topological and trivial cases, respectively.

Using the above RSTIs, we find that, for the process with $\Phi = k_z$ varying from 0 to $2\pi$, the imbalance of the bulk eigenstates in the topological case is given by $\Delta(g_0) = m(g_2) -$

$m(g_0) = \delta_1 = -1$, $\Delta(g_1) = m(g_0) - m(g_1) = -\delta_2 = 1$ and $\Delta(g_2) = m(g_1) - m(g_2) = \delta_2 - \delta_1 = 0$. This implies that there is one excess bulk state with the IR $A$ below the band gap and two excess bulk states with the IRs $^2E$ and $^1E$ above the band gap. The nonzero $\Delta(g_0)$ and $\Delta(g_1)$ account for the cyclic transformation of excess bulk states that cannot find their partners within the bulk bands they belonged to. The resultant spectral flow across the band gap is presented in Fig 4(a). Numerical calculations that indicate the robustness of the spectral flows against various disorders are discussed in Supplementary Material.

We now turn to the analysis of the SSH model. We consider a semifinite system with $2N$ unit-cells along the $x$-direction (periodic in $y$- and $z$-directions), as depicted in Fig. 3(d). This finite system has $C_2$ rotation symmetry and can thus be divided into two sectors related to each other under $C_2$. After the Fourier transformation for both $y$- and $z$-directions, there are $2N - 1$ bulk states in each bulk band in the finite system in the topological phase (we consider the system a specific $k_y$). For comparison, in the trivial phase, there are $2N$ bulk states in each bulk band. In the topological phase, there are two edge states in the bulk band gap.

Because of the strange number of bulk states below the band gap, each sector has a fractional charge of $e/2$ if the bulk states below the band gap are all filled, because the filling of the first bulk band yields $N - \frac{1}{2}$ bulk states in each sector. Notice that since there are two edge states in the bulk band gap, whether they are filled or empty, they do not affect the fractional charge. This fractional charge is known for decades but observed only recently [56].

The strange number of the bulk states in each band is also reflected by the topological Wannier cycles. Inserting a gauge flux at the center of the SSH model leads to the cyclic transformation between the states of different IRs, $A$ and $B$ (or equivalently, $g_0^{(2)}$ and $g_1^{(2)}$), i.e., $g_0^{(2)} \to g_1^{(2)} \to g_0^{(2)}$. Since each band has $2N - 1$ bulk states, there is one bulk state having no partner of transformation within the bulk band it belongs to. Therefore, there must be a cyclic transformation formed by two states belonging to different bulk bands. Such cyclic transformation leads to the topological Wannier cycle which is manifested as the spectral flows across the band gap. This topological Wannier cycle is realized by the glide step dislocation in Fig. 3(b). The energy spectrum that confirms the emergent topological Wannier cycle explicitly is given in Fig. 4(b) which is obtained from the calculation based on the TB model in Fig. 3(d). We emphasize that, interestingly, here the topological Wannier cycles are manifested as the 1D edge states in 2D SSH-like models (with a special zigzag edge boundary). This provides a new insight and an unprecedented mechanism for gapless edge states in 2D topological crystalline systems.

The RSTI for our $C_2$-symmetric SSH model is defined as a single number [39],

$$\delta = m(g_1) - m(g_0). \tag{3}$$

Similarly, it can be calculated as follows in terms of the IRs at HSPs in momentum space [39],

$$\delta = -\frac{1}{2}m(A_\Gamma) - \frac{1}{2}m(B_X). \tag{4}$$

We find $\delta = -1$ for the topological band gap. It implies that the states with $C_2$-rotation eigenvalues $g_0$ and $g_1$ redundant in two different bands transform cyclically in the band gap. For the trivial band gap, we find $\delta = 0$ and thus no spectral flow emerges.

The above analysis for both the breathing kagome model and the SSH model establishes a firm connection between the filling anomaly (or equivalently the fractional charge) and the topological Wannier cycles. We emphasize that unlike the corner states protected by the chiral symmetry which are commonly used as a signature of higher-order topology, the filling anomaly does not rely on the chiral symmetry. In fact, breaking chiral symmetry does not affect the above analysis and reasoning. Thus, the topological Wannier cycles are robust against chiral symmetry breaking. As shown in a recent experiment [34], topological Wannier cycles can emerge in phononic systems where the chiral symmetry is broken. Intriguingly, in such phononic systems, chiral symmetry breaking destroys the edge and corner states, yet the topological Wannier cycles remain as a salient spectral feature with robust, gapless topological boundary states [34]. We also present the phononic realization of the above studied models in Supplementary Material. The numerical calculation agrees well with the above analysis.

## Coexisting bulk and edge topological Wannier cycles

It is noticed that the topological Wannier cycles studied above and before only involve bulk states, i.e., the cyclic spectral flows start from and end at the bulk continua. Here, taking a step forward, we extend the physics of topological Wannier cycles to surface states (or edge states). One example is illustrated in the TB model in Fig. 5(a), where we create a step glide dislocation in the $C_4$-symmetric 2D SSH model with four sites in a unit-cell. In the topological phase when the intra-unit-cell coupling $|t_1|$ is weaker than the inter-unit-cell coupling $|t_2|$, the Wannier centers are at the corner of each unit-cell [see the blue dots in Fig. 5(a)]. However, some Wannier centers are exposed at the edge regions in the finite system [see the purple dots in Fig. 5(a)], which leads to the emergence of edge states in the two band gaps. Intriguingly, the gauge flux introduced by the step glide dislocation acts on the Wannier orbitals not only in the bulk region but also the upper and lower edges [see the orange zone in Fig. 5(a)], which results in the topological Wannier cycles of both bulk and edge states. Such simultaneous Wannier cycles are verified in the energy spectra as functions of the artificial gauge flux $\Phi = k_z$ in Fig. 5(b). The spectral flows of edge states are denoted by red curves, which are analogous to those of bulk states in Fig. 4(b) but are confined in the 2D edges in the *x-z* planes. Interestingly, these phenomena are manifested as the coexisting 2D topological modes bound to the interface and 1D topological modes bound to the edge of the interface. Both of them are gapless but sharing distinct spectral properties. The bulk topological Wannier cycles (they are many due to the many bulk Wannier centers encircled by local gauge fluxes) traversing the two bulk band gaps, whereas the edge topological Wannier cycles traversing the edge band gap [Fig. 5(b)]. The intriguing phenomenon of edge topological Wannier cycles is further confirmed by studying the topological Wannier cycles for the edges of the quadrupole topological insulators [see Figs. 5(c) and 5(d)]. Here, as the bulk topological Wannier cycles disappear (since the quadrupole topological insulators are not Wannier representable), only the edge topological Wannier cycles emerge (see Supplemental Material for more details).

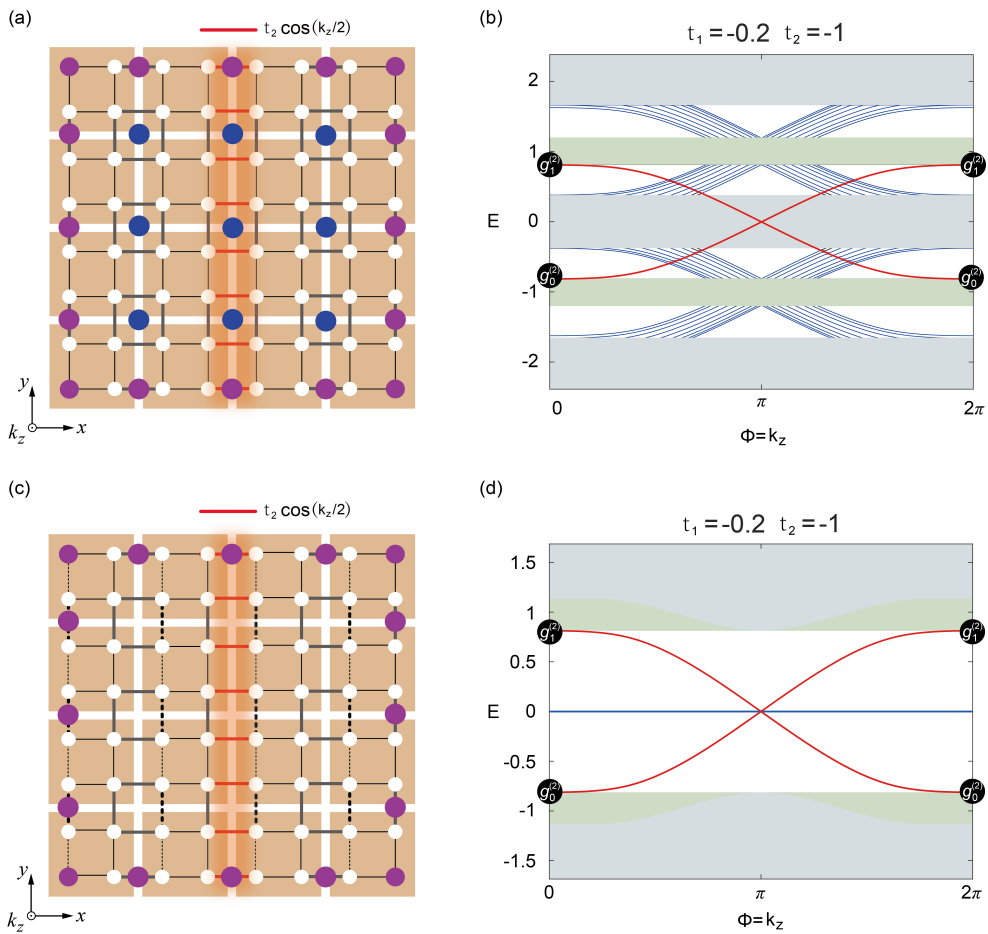

Fig. 5 | (Color online) Topological Wannier cycles generalized to edge states. (a) The effective TB 2D SSH model with a step glide dislocation. The solid thin and thick lines denote the intra- and inter-unit-cell couplings $t_1$ and $t_2$, respectively. The red lines represent the couplings $t_2 \cos\left(\frac{k_z}{2}\right)$, which bring in the gauge flux $k_z$, as sketched by the orange zone. The blue (purple) dots denote the Wannier centers at the bulk (edge)

region. (b) The corresponding energy spectra in (a) as functions of the gauge flux $k_z$. The grey and green regions denote the bulk and edge states, respectively. The spectral flows of bulk and edge states are marked with blue and red colors, respectively. The TB parameters are given on top of each panel. (c) The effective TB 2D quadrupole model with a step glide dislocation. The difference compared to the 2D SSH model in (a) are the negative hoppings -$t_1$ and -$t_2$, denoted by the dashed thin and thick lines, respectively, which makes the bulk states not Wannier representable. (d) The eigen-energy spectra of the system in (c) as functions of the gauge flux $k_z$. The grey and green regions denote the bulk and edge states, respectively. The spectral flows of edge states are marked with red colors. The blue line denotes the zero-energy corner states. The The TB parameters are given on top of each panel.

## Conclusion and outlook

In this work, we generalize the study on topological Wannier cycles to systems with various spatial symmetries, revealing rich way to create highly localized artificial gauge flux and the topological modes bound to the gauge flux in crystalline systems. In particular, for systems with *C₂* symmetry, we uncover an unprecedented mechanism for gapless edge states through topological Wannier cycles with the zigzag edge boundaries. We further extend topological Wannier cycles to systems with both bulk and edge states, and show simultaneous emergence of bulk and edge topological Wannier cycles in 3D systems that are finite in two dimensions, revealing rich topological phenomena induced by local artificial gauge flux in crystalline systems. Our findings can also be extended to electronic systems where fractionalization may also include the spin degree of freedom, and rich topological Wannier cycles are expected that may be induced by screw dislocations and glided edge boundaries in solid state materials.